\documentclass[preprint]{ptephy_v1}

\preprintnumber{XXXX-XXXX} 
\usepackage{hyperref}

\usepackage{amsmath} 
\usepackage{amsthm} 
\usepackage{hyperref} 
\usepackage{graphics} 
\usepackage{url}
\usepackage{dirtytalk}
\usepackage{xspace}
\usepackage{bm}

\newcommand{\ts}{\textsuperscript}
\newcommand{\gammaray}{\ensuremath{\gamma}-ray\xspace}
\newcommand{\gammarays}{\ensuremath{\gamma}-rays\xspace}
\newcommand{\etwop}{\ensuremath{E(2^+_1)}\xspace}
\newcommand{\efourp}{\ensuremath{E(4^+_1)}\xspace}
\newcommand{\etwopt}{\ensuremath{2^+_1 \rightarrow 0^{+}_{\mathrm{gs}}}\xspace}
\newcommand{\efourpt}{\ensuremath{4^+_1 \rightarrow 2^+_1}\xspace}

\newcommand{\stwop}{\ensuremath{2^+_1}\xspace}
\newcommand{\sfourp}{\ensuremath{4^+_1}\xspace}

\newcommand{\beupl}{\ensuremath{B(E2;~0^+_{\mathrm{gs}}\rightarrow2^+_1)}\xspace} 
\newcommand{\ppp}{\ensuremath{(p,p') } }

\newcommand{\dali}{DALI2\ts{+}\xspace}
\newcommand{\ioi}{Island of Inversion\xspace}
\newcommand{\iois}{IoI\xspace}

\begin{document}

\title{The Island of Inversion at N=40}

\author[]{Martha Liliana Cort\'es}
\affil[]{RIKEN Nishina Center, Wako, Saitama 351-0198, Japan}




\begin{abstract}%
Our understanding of the structure of atomic nuclei largely derives from the nuclear shell model, which has proven widely successful. 
Further test to our interpretation of the nuclear properties is provided by the study of shell evolution.  Increasing experimental information has shown that the nuclear energy shells change when going towards the most exotic nuclei, in turn making some shell closures disappear while others arise. 
In particular, the $N=40$ sub-shell closure has been the subject of extensive research due to the emergence of a so-called \ioi, where deformed intruder configurations dominate the wave function of the ground state. An overview of recent experimental results in the $N=40$ \ioi, particularly those performed with the combination of the MINOS hydrogen target and the DALI2 \gammaray array at the RIBF are discussed. 
\end{abstract}

\subjectindex{xxxx, xxx}

\maketitle


\section{Introduction}



The concept of nuclear shell structure plays a key role in our description of the atomic nucleus.  
In this model, nucleons  occupy discrete energy levels determined by the effective central potential.
Sizable gaps between groups of energy levels disfavour the excitation of particles to higher lying orbitals, giving rise to the so-called magic numbers. 
A complete understanding of the magic numbers experimentally observed in stable isotopes (2, 8, 20, 28, 50, 82, 126) requires to consider not only a harmonic oscillator-type central potential, but also to include the spin-orbit interaction~\cite{Goeppert_PhysRev.75.1969,Haxel_PhysRev.75.1766.2}. 


In the last decade an increasing amount of experimental information has been gathered which indicates that the nuclear shell closures are modified when going away from the valley of stability~\cite{Sorlin_PPNP_2008,Otsuka_RevModPhys.92.015002_2020, Nowacki_PPNP_120_2021}. 
A particular interesting example is the disappearance of the $N=20$ magic number and the formation of a so-called Island of Inversion (IoI) around \ts{32}Mg, where deformed intruder configurations become the ground state. 
Different theoretical interpretations have been developed, and it has been identified that Islands of Inversion are formed
due to  quadrupole correlations  which maximize energy gains with multiparticle-multihole excitations across the gap at the magic numbers~\cite{Nowacki_PPNP_120_2021}.

Another region of the nuclear chart that has attracted much interest due to the development of an \iois is the one of the neutron-rich isotopes around $N=40$, which corresponds to a shell closure of the harmonic oscillator potential. 
Investigations on the possible magicity of \ts{68}Ni have evidenced a high \etwop energy~\cite{Broda_PhysRevLett.74.868_1995} and a relatively small \beupl value~\cite{Sorlin_PhysRevLett.88.092501_2002,Bree_PhysRevC.78.047301_2008}, suggesting a large energy gap between between the $pf$ shell and the $g_{9/2}$ orbital. 
However, the neutron separation energy shows no hints of magicity, which  has prompted the suggestion that the low \beupl value arises from a neutron dominated excitation and not from a shell closure~\cite{Langanke_PhysRevC.67.044314_2003}. Furthermore, well deformed structures have been observed at low excitations energies in \ts{68}Ni which become the ground state in Fe and Cr isotopes. 

Further studies on the $N=40$ \iois below \ts{68}Ni, as well as its possible extension towards $N=50$,  have been recently performed at the  Radioactive Isotope Beam Factory (RIBF) of the RIKEN Nishina Center in Japan, largely with the use of the MINOS liquid hydrogen target~\cite{Obertelli_EPJA_50_2014} and the DALI2~\cite{Takeuchi_NIMA763_2014} and \dali~\cite{Murray_RAPR_2018} arrays for \gammaray detection.  In this article, we review some recent experimental results on even-$N$ isotopes around the $N=40$ \iois, in particular using direct reactions with MINOS.  

Section~\ref{sec:theory} presents the theoretical framework employed, while Sec.~\ref{sec:experiment} describes the experimental setup. 
Results on even- and odd-$Z$  isotopes are presented in Sec.~\ref{sec:even} and \ref{sec:odd}, respectively. Future perspectives on the region are discussed in Sec.~\ref{sec:perspectives}.  
More detailed reviews on shell evolution and  the $N=20$ and $N=40$ \ioi can be found in Refs.\cite{Nowacki_PPNP_120_2021,Otsuka_RevModPhys.92.015002_2020,Gade_physics3040077}.


\section{Theoretical approach}\label{sec:theory}

A remarkable shell  model calculation which has been successfully used to explain most of the observable in this region is the one based on the Lenzi-Nowacki-Poves-Sieja (LNPS) interaction, which was published in 2010, ahead of many of the experimental results~\cite{Lenzi_PhysRevC.82.054301}.

Previous theoretical studies reproduced early experimental information in the region, such as the level schemes of \ts{62,64}Fe, by using a realistic interaction and a valence space including the $pf$ shells for protons and the  $0f_{5/2}$, $1p_{3/2}$, $1p_{1/2}$, and $0g_{9/2}$ orbits for neutrons~\cite{Sorlin_EPJA_2003}.
However, such calculations were not successful in making predictions towards heavier isotopes, with the necessity of including the $2d_{5/2}$ orbital  pointed out~\cite{Caurier_EPJA_2002}.
This necessity arises from the fact that in the inclusion of the quasi-SU(3) sequence $0g_{9/2}$, $1d_{5/2}$,  $2s_{1/2}$ together with the pseudo-SU(3) triplet $1p_{3/2}$, $0f_{5/2}$, $1p_{1/2}$  allows the development of quadrupole correlations which results in deformation~\cite{Nowacki_PPNP_120_2021}. 

The LNPS interaction is based on  realistic two-body matrix elements with experimental constrains used to tune the monopole Hamiltonian. A model space based on a \ts{48}Ca core, the $pf$ shell for protons and the $1p_{3/2}$, $1p_{1/2}$, $0f_{5/2}$, $0g_{9/2}$, and $1d_{5/2}$ orbits  for neutrons is adopted. 
As the third member of the quasi-SU(3) sequence, $2s_{1/2}$, is not included, its effect is compensated by an increase of the quadrupole-quadrupole interaction of the $0g_{9/2}$, and $1d_{5/2}$ orbits.
Minor modifications to the monopole and pairing parts of the Hamiltonian have been included along the years to allow the description of a broader region. Although in some publications such modified interaction is referred as LNPS-m or LNPS-new, within this document the term LNPS will be used to refer to any version of the original interaction. 

\begin{figure}[bth]
    \centering
    \includegraphics[width=0.6\textwidth]{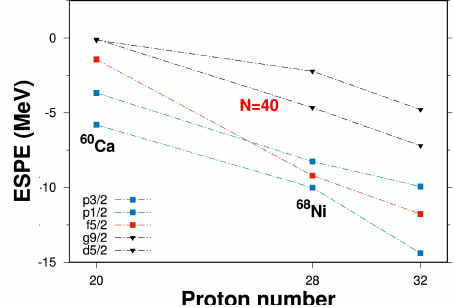}
    \caption{Neutron effective single particle energies at $N=40$ from the LNPS interaction. Reprinted from~\cite{Nowacki_PPNP_120_2021} with permission from ELSEVIER. }
    \label{fig:ShellModel}
\end{figure}

Figure~\ref{fig:ShellModel} shows the neutron effective single particle energies (ESPE) of the LNPS interaction where a reduction of the $0f_{5/2}-0g_{9/2}$ gap  below \ts{68}Ni is observed. Furthermore it can be seen that there is a very narrow separation between the $0g_{9/2}$ and the $1d_{5/2}$ orbits. 

The calculation predicts a maximum on the deformation for \ts{64}Cr, with the lowest \etwop and the highest \beupl value.
For \ts{68}Ni the wave function contains a large component corresponding to spherical configuration, although  the occupation of the $0g_{9/2}$ is already visible and keeps increasing, as well as the occupation of the $1d_{5/2}$, when protons are removed. 
Regarding the ground state wave function, the 4p-4h component is dominant for Fe, Cr and Ti, although the 2p-2h and 6p-6h components are not negligible. This result is in contrast to the case of the $N=20$ \iois, where the 2p-2h components dominates the ground state wave function. 
For \ts{60}Ca, the LNPS calculation predicts a  \etwop  of around 1.5~MeV.  This increase in energy as compared to Ti, however, does not imply a double-magic character. The ground state wave function has more than three neutrons above the $N=40$ gap, and only a 5\% component from the doubly-magic configuration which place it within the \iois~\cite{Lenzi_PhysRevC.82.054301,Nowacki_PhysRevLett.117.272501}.

The shell model calculations using the LNPS interaction provide a consisten framework to study the formation of an \iois at $N=40$, which extends down to \ts{60}Ca. It is noted that this situation,  where deformed intruder states become energetically favoured, is described as  due to the interplay between 
the presence of valence protons in the normal configuration and the reduction of the $N=40$ gap approaching $Z=20$. 


\section{Experimental approach}\label{sec:experiment}

In recent years quasi-free scattering (QFS) and knockout reactions in inverse kinematics have been the preferred tools to measure isotopes very far from the valley of stability.
Such direct reactions have the advantage of a high selectivity to probe  individual nuclear orbitals, and offer the possibility to obtain information on angular momentum of the removed nucleon  by  measuring the momentum distribution of the residual fragment. 

Exotic  $N=40$ isotones below \ts{68}Ni have been measured at different facilities  using  a plethora of techniques~\cite{Nowacki_PPNP_120_2021,Gade_physics3040077}. However, due to the low production cross section, reaching the most exotic species requires an increasingly high beam intensity.
In the last decade, the combination of the highest primary beam intensities provided by the RIBF, with the MINOS liquid hydrogen target~\cite{Obertelli_EPJA_50_2014} has provided a significant amount of new information on exotic nuclei. In particular, by surrounding MINOS with the DALI2 and DALI2\ts{+} arays~\cite{Takeuchi_NIMA763_2014,Murray_RAPR_2018} the structure of the most exotic neutron-rich isotopes could be investigated within the SEASTAR project~\cite{seastar_web}.

Isotopes around  the  $N=40$ \iois  were produced during the SEASTAR~1 and SEASTAR~3 campaigns carried out in 2014 and 2017, respectively. Primary beams of \ts{238}U and \ts{70}Zn, with intensities of 12 and 240~pnA, respectively were used to produce the isotopes of interest, which were separated and identified with the BigRIPS separator~\cite{Kubo_PTEP2012_2012}.  
For the case of SEASTAR~1  outgoing fragments were identified with the ZeroDegree spectromenter~\cite{Kubo_PTEP2012_2012} and \gammarays were detected with the DALI2 array composed of 186 NaI(Tl) detectors~\cite{Takeuchi_NIMA763_2014}, while for SEASTAR~3, the SAMURAI magnet~\cite{Kobayashi_NIMB_317_2013} was used for the outgoing particle identification and the upgraded DALI2\ts{+} array consisting of 226 NaI(Tl) crystals~\cite{Murray_RAPR_2018} was used for \gammaray detection. 

The use of the liquid hydrogen target of MINOS allowed to increase the luminosity and to  reduce the background for secondary reactions. Target thicknesses of 100 and 150~mm were used for SEASTAR~1 and  SEASTAR~3, respectively. The use of the time projection chamber which surrounds the target is indispensable to reconstruct the reaction-vertex position and improve the energy resolution obtained after Doppler correction. For the case of the MINOS target, it was possible to reconstruct the vertex position with a resolution of 5~mm FWHM~\cite{Santamaria_PhysRevLett.115.192501}.


\section{The even$-Z$ isotopes}\label{sec:even}


\subsection{Fe isotopes}

\begin{figure}[bt]
    \centering
    \includegraphics[angle=-90,width=\textwidth]{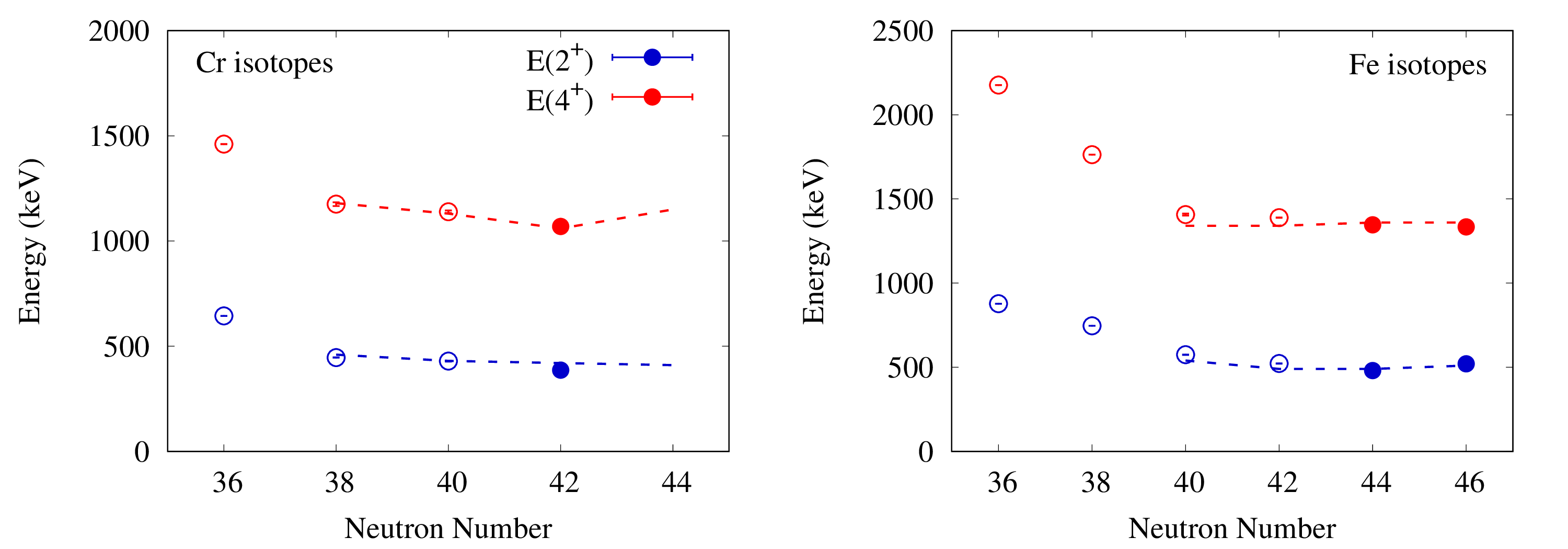}
    \caption{Systematics of experimental \etwop and \efourp  for Cr (left ) and Fe (right)  isotopes. Open circles represent previous literature values, while the filled squares indicate the results obtained within the SEASTAR project. Theoretical calculations with the LNPS interaction are  shown by the dashed lines. Values taken from Ref.~\cite{Santamaria_PhysRevLett.115.192501}.}
    \label{fig:isotopesE}
\end{figure}

As displayed in Fig.~\ref{fig:isotopesE}, the  \etwop and \efourp  of \ts{64,66}Fe ($N=38, 40$) show a significant decrease as compared to lighter Fe isotopes, indicating an increase of collectivity when approaching $N=40$~\cite{Hannawald_PhysRevLett.82.1391_1999}. 
Such conclusion has been supported by the measured transition probabilities in \ts{66}Fe~\cite{Rother_PhysRevLett.106.022502,Crawford_PhysRevLett.110.242701}.
Going beyond $N=40$, the \etwop  of \ts{68}Fe is only slightly reduced as compared to \ts{66}Fe~\cite{Adrich_PhysRevC.77.054306_2008}, while a modest increase on the \beupl was reported~\cite{Crawford_PhysRevLett.110.242701}. These results have been interpreted  by shell model calculations with the LNPS interaction, which reproduce the data and indicate the importance of excitations to the $g_{9/2}$ and $d_{5/2}$ orbitals in the understanding of the structure of isotopes in the $N=40$ \iois.

The first spectroscopy of \ts{70}Fe has been measured at the  RIBF  via $\beta-$decay~\cite{Benzoni_PLB_751_2015} as well as in-beam \gammaray spectroscopy~\cite{Santamaria_PhysRevLett.115.192501}.  The latter measurement, which was the first published result of the SEASTAR project, also obtained the first spectroscopy of \ts{72}Fe. 
In the experiment, excited states of \ts{70,72}Fe and \ts{66}Cr were populated by $(p,2p)$ and $(p,3p)$ reactions in MINOS. Two transitions were clearly visible in the DALI2 spectra corresponding to each isotope and, assigned to the decay of the \stwop and \sfourp states based on the measured intensities, coincidence analysis, and systematic. 
The results are displayed in Fig.~\ref{fig:isotopesE}, where it can be seen that the measured \etwop and \efourp  along the Fe isotopes show a seemingly constant value between $N=40$ and $N=46$, suggesting that there is no further increase of the deformation beyond $N=40$ and that these isotopes remain within the $N=40$ \iois and extend it towards $N=50$. 

Shell model calculations employing the LNPS interaction in the $pf$ model space for protons and the $p_{1/2}f_{5/2}g_{9/2}d_{5/2}$ for neutrons  reproduce the measurements successfully, as shown by the dashed lines in Fig.~\ref{fig:isotopesE}, and suggest that the neutron-rich Fe isotopes  behave as prolate deformed rotational nuclei. 

More recent spectroscopic measurements on \ts{70}Fe using the high-resolution Ge array  GRETINA~\cite{Paschalis_NIMA_2013,Weisshaar_NIMA_2017} following proton removal reactions, have reported in addition a candidate for the $6^+$ state at 2444~keV, as well as effective lifetimes of $\tau=120(20)$~ps and $\tau=2.3(1.5)$~ps for the \stwop and \sfourp states, respectively~\cite{Gade_PhysRevC.99.011301}. 
Although these lifetime values represent an upper limit due to possible unobserved feeding, a general good agreement  with the LNPS calculations is observed. Furthermore, the calculations place the $6^+$ state within 30~keV of the experimental value, showing their remarkable predictive capabilities.

\subsection{Cr isotopes}

In a similar way as for the Fe isotopes, for the Cr isotopic chain a strong deformation when reaching $N=40$ was suggested. These conclusions were based on previous measurements of the \etwop  of \ts{62,64}Cr ($N=38,40$)~\cite{Sorlin_EPJA_2003, Gade_PhysRevC.81.051304}, and  supported by the measurements of transition probabilities~\cite{Aoi_PhysRevLett.102.012502_2009,Baugher_PhysRevC.86.011305_2012}.
Furthermore, a recent study on \ts{62}Cr reported for the first time  an excited $0^+$ state, revealing shape coexistence in this region~\cite{Gade_NaturePhys_2024}. 

Beyond $N=40$, \ts{66}Cr is the heaviest Cr isotope for which spectroscopic information is available. The measurement was performed within the SEASTAR project, together with the Fe isotopes as described above. Two transitions were observed in the \gammaray spectrum measured with DALI2 following quasi-free proton knockout, and assigned to the \stwop and \sfourp states of \ts{66}Cr. Figure~\ref{fig:isotopesE} displays the obtained results together with the systematics of \etwop and \efourp  of lighter Cr isotopes. It can be seen that the Cr isotopes follow a similar trend as the Fe isotopes with the \etwop and \efourp  remaining  constant from $N=38$ to $N=42$, and with \ts{66}Cr showing the lowest \etwop  in the region at 386~keV, slightly below the value of \ts{64}Cr of 429~keV. 

Shell model  calculation using the LNPS interaction reproduce the experimental data, as shown by the dashed lines in Fig.~\ref{fig:isotopesE}. The calculations show that the deformation stays constant, with a maximum at $N=40$ where an intrinsic shape with deformation parameter $\beta\sim0.33$ arises resulting in the constant energy values. Despite this, a varying degree of particle-hole excitations across $N=40$ is predicted along the isotopic chain~\cite{Santamaria_PhysRevLett.115.192501}. 

Further shell model calculations considering the $pf$ shell for protons and the $sdg$ shell for neutrons and a \ts{60}Ca core, have been performed to study the evolution of the $N=40$ \ioi towards $N=50$.  An effective interaction containing some modifications to the LNPS interaction and termed PFSDG-U has been employed~\cite{Nowacki_PhysRevLett.117.272501}.
Such  calculations also  highlight the interplay between quadrupole and pairing correlations and predict a merging of the  $N=40$ \iois with a new one formed around $N=50$, in a similar way as the $N=20$ \iois continues towards $N=28$.


\subsection{Ti isotopes}
Experimental information for  Ti isotopes around $N=40$ is limited. 
Early measurements on  production cross sections from the fragmentation of a \ts{76}Ge beam indicated that the $Z=19-22$ neutron-rich nuclei display an enhanced cross section with respect to the lighter isotopes and turned to be  more  bound than predicted by  mass models~\cite{Tarasov_PhysRevLett.102.142501}.

The first spectroscopy of \ts{60}Ti, measured at NSCL with the GRETINA array, showed \etwopt and \efourpt transition energies of 850(5) and 866(5)~keV, respectively~\cite{Gade_PhysRevLett.112.112503}. Shell model calculations both with and without including the  $g_{9/2}$ orbital were able to reproduce the observed excited states. However,  only the calculation  including excitations above $N=40$ could successfully reproduce the measured knockout cross sections.  

Isomer spectroscopy has revealed two new isomers in \ts{61}Ti~\cite{Wimmer_PLB_792_2019}.
Comparison of this  result with shell model calculations based on the LNPS interaction suggested that particle-hole excitations across $N=40$ dominate the ground state configurations of \ts{59,61} Ti, placing them  within the $N=40$ \iois.

Recent mass measurements on Sc, V and Ti isotopes using the time-of-flight method show that the two-neutron separation energy, $S_{2n}$,  is flat for the Ti isotopes when going towards $N=40$, indicating that they are more stable than expected~\cite{Michimasa_PhysRevLett.125.122501}.



\begin{figure}[b!]
    \centering
    \includegraphics[angle=-90,width=0.55\textwidth]{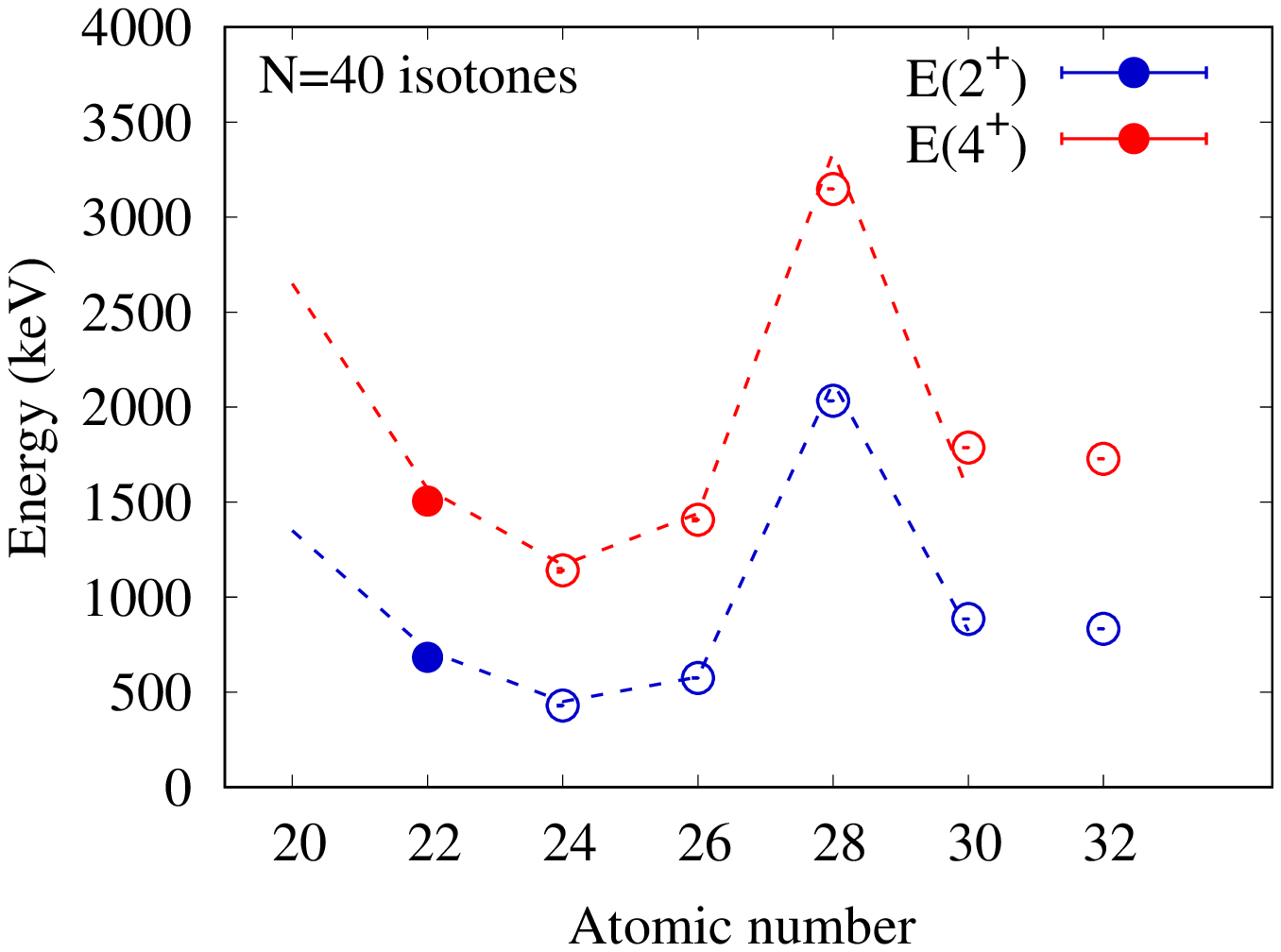}
    \caption{\etwop and \efourp  of even-even $N=40$ isotones. Available experimental data is compared with the results of shell model calculations using the LNPS interaction, shown by the dashed lines. The filled circles indicate the results obtained within the SEASTAR project. Values taken from Ref.~\cite{Cortes_PhysLettB.800.135071}.}
    \label{fig:isotonesE}
\end{figure}

Obtaining spectroscopic information along the $N=40$ isotones when going from \ts{64}Cr to the lighter even-$Z$ isotope, namely \ts{62}Ti, took 10 years and could only be achieved at the RIBF within the SEASTAR project. 
In the experiment, \ts{62}Ti was populated by proton knockout reactions from \ts{63}V in the MINOS target. \gammaray spectroscopy performed with the DALI2\ts{+} array showed two transitions at 683(10) and 823(20)~keV tentatively assigned to the decays of the  \stwop and \sfourp states, respectively~\cite{Cortes_PhysLettB.800.135071}. 
Comparison with theoretical predictions based on the  LNPS interaction, performed already in 2010, is exceptionally good as can be seen in Fig.~\ref{fig:isotonesE} where the known experimental  \etwop and \efourp of even-even  $N=40$ isotones are compared to these calculations. 
Within this theoretical framework the ground state of \ts{62}Ti is dominated by 4p-4h excitations with a significant component of 6p-6h resulting from a gap of about 1~MeV between the $f_{5/2}$ and the $g_{9/2}$. This result places \ts{62}Ti within the $N=40$ \iois and suggests that it extends to \ts{60}Ca.


\subsection{Ca isotopes}

For the Ca isotopic chain towards $N=40$, spectroscopic information of \ts{56,58}Ca, beyond the proposed $N=34$ shell closure, has been measured for the first time within the SEASTAR project by proton knockout reactions off \ts{57,59}Sc in the MINOS target. In \ts{56}Ca a \gammaray transition was detected at 1456(12)~keV, while in \ts{58}Ca,  a possible transition at 1115(34)~keV was observed with a significance of 2.8$\sigma$. Both transitions were attributed to the decay of the \stwop state in the corresponding Ca isotopes~\cite{Chen_PLB_2023}. 

For Ca isotopes up to \ts{54}Ca ($N = 34$), a good agreement has been obtained with shell model calculations employing the GXPF1 family of interactions~\cite{Honma_EPJA_25_2005, Steppenbeck_nature_2013, Chen_PhysRevLett.123.142501}. In these calculations, the consideration of the full $pf$ shell is sufficient to reproduce the energies of lying states, and reproduce the \etwop  at the $N=32$ and $N=34$ sub-shell closures. Within the same model, the \etwop  of \ts{56,58}Ca should remain constant.
The experimentally observed reduction of excitation energy when going from $N=36$ to $N=38$ suggest that the $f_{5/2}$  does not behave like an isolated orbital but instead couples to the $g_{9/2}$ orbital, preventing the formation of an $N=40$ gap around \ts{60}Ca. 
Shell model calculations with the A3DA-m interaction, fitted to reproduce the experimental data, suggest a pairing between these two orbitals with a consequential prediction of the drip line in Ca isotopes at $N=50$~\cite{Tsunoda_PhysRevC.89.031301_2014, Chen_PLB_2023}.

Spectroscopic information on odd-$A$ Ca isotopes was also obtained within the SEASTAR project. 
The measurement of a short-lived state in \ts{57}Ca at 751(13)~keV suggests a structural change at $N=36$ where the particle-hole configurations are dominant, in agreement with the development of the $N=40$ \ioi~\cite{Koiwai_PLB_2022}. Further information on the structure of the Ca isotopes recently obtained at the RIBF can be found elsewhere~\cite{Ca_PTEP}.

The discovery of the $N=40$ isotope of Ca, namely \ts{60}Ca, from the fragmentation of a \ts{70}Zn beam at the RIBF has been only recently reported~\cite{Tarasov_PhysRevLett.121.022501}.
Although spectroscopy of \ts{60}Ca cannot be achieved at the existing facilities, it will become accesible after the planned  upgrade of the RIBF~\cite{ribf_update}.
Based on the good agreement between the different experimental measurements, in particular the spectroscopy of \ts{62}Ti, with the LNPS calculations, an \etwop  or around 1.5~MeV is to be expected, followed by a triplet of $0^+$, $2^+$ and $4^+$ states~\cite{Nowacki_PPNP_120_2021}. 


\section{The odd$-Z$ isotopes}\label{sec:odd}

\subsection{Co isotopes}

The Co isotopic chain, with only one proton hole below the magic $Z=28$, has been the subject of extensive research as nuclei with only one particle or hole outside of a magic shell closure allow to study the interplay between the collective and single-particle degrees of freedom, and to investigate  the strength of the shell gaps. 
Indeed, the structure of Co isotopes up to $N=38$  is well described by the coupling of the $\pi f_{7/2}^{-1}$ proton hole to its adjacent Ni isotope, asserting the conservation of the $Z=28$ gap. 

The discovery of a $1/2^-$ isomeric state in \ts{67}Co with a surprisingly low energy of 492~keV~\cite{Pauwels_PhysRevC.78.041307} was interpreted within the Nilsson model as a collective structure at $N=40$ arising from an intruder state.  Furthermore, the $3/2^-$ and $5/2^-$ states were suggested to belong to a rotational band on top of this deformed structure. 
On the other hand, the measured $9/2^-$ and $11/2^-$ states followed the systematic of lighter Co isotopes and could be interpreted as the core-coupled states. 
Such a discovery pointed  at shape coexistence when reaching the  Ni isotopes. 
Further information on the structure towards \ts{78}Ni, studied within the SEASTAR project, can be find elsewhere~\cite{Ni_PTEP}.

The structure of \ts{65}Co, at $N=38$, showed similar features, 
which was interpreted as arising from the influence of the deformed intruder configuration, although not as well developed as in \ts{67}Co~\cite{Pauwels_PhysRevC.79.044309}. 
By using shell model calculations with large model spaces and, in particular, the LNPS interaction, a simultaneous description of both the core-coupled and the deformed structures up to \ts{67}Co was possible, showing the importance of including the $g_{9/2}$ and $d_{5/2}$ orbitals to account for the deformed configurations~\cite{Recchia_PhysRevC.85.064305_2012}.

For the case of \ts{69}Co, $\beta-$decay studies have reported few \gammaray transition, which, however, could not be placed in a level scheme~\cite{Liddick_PhysRevC.92.024319_2015}. The presence of a 750(250)~ms $\beta-$decaying isomer was also reported and  suggested to correspond to the $1/2^-$ deformed state similar to the ones observed in \ts{65,67}Co. Only recently the excitation energy of this isomer has been reported to be 176(57)~keV, indicating the persistence of shape coexistence in the region~\cite{Canete_PhysRevC.101.041304_2020}.

\begin{figure}[b!]
    \centering
    \includegraphics[angle=-90,width=\textwidth]{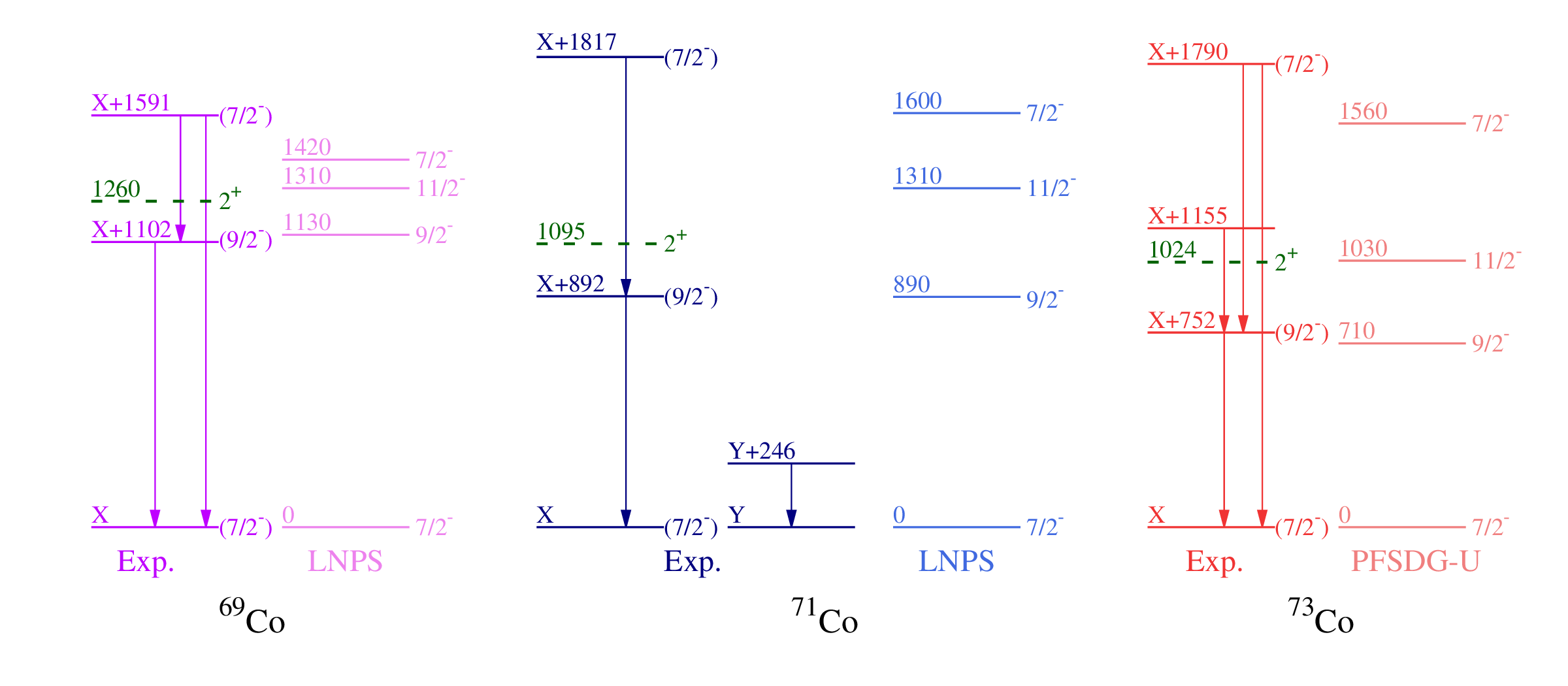}
    \caption{Experimental level schemes of \ts{69,71,73}Co and comparison with theoretical calculations with the LNPS and PFSDG-U interactions. For comparison the \stwop state of the corresponding Ni isotopes are displayed by the green dashed lines. Values taken from~\cite{Lokotko_PhysRevC.101.034314}}
    \label{fig:CoEvolution}
\end{figure}

Spectroscopic information of Co isotopes beyond $N=40$ has been accessible within the SEASTAR project. Excited states in \ts{69,71,73}Co were populated  by proton knockout reactions on \ts{70,72,74}Ni and their de-excitation \gammarays measured with the DALI2 array~\cite{Lokotko_PhysRevC.101.034314}. 
Level schemes for the three isotopes were proposed based on the $\gamma-\gamma$ coincidence analysis.
In order to assign the spin of the measured levels, shell model calculations were performed. 
For \ts{69,71}Co, the $pf_{5/2}g_{9/2}d_{5/2}$ model space together the LNPS interaction was employed, while for \ts{73}Co the $pf-sdg$ model space and the PFSDG-U interaction was used. 
Based on the agreement of the calculations with the experimental level schemes, candidates for the $9/2^-$ and $7/2^-$ states in each isotope were proposed. 
Figure~\ref{fig:CoEvolution} shows the experimental level schemes as well as the theoretical calculations for \ts{69,71,73}Co.
As a comparison, the  \etwop  of the neighboring  Ni isotopes are also displayed as a green dashed line.  It can be seen that the $9/2^-$ state follows the trend of the \etwop, suggesting that core coupling is maintained in this region. 

The calculations  predict that, in addition to this spherical structure based  on the  $\pi f_{7/2}^{-1} \bigotimes 2^+$(Ni), a coexisting  band with a deformed  structure based on a $\pi f_{7/2}^{-2} pf^{+1}$ configuration exists, similarly as for the lighter Co isotopes. 
Although low energy transitions were measured during the SEASTAR experiment, the placement of such a deformed structure could not be determined. 

Interestingly, the measurement of the $\beta-$decay of \ts{75}Co performed at the RIBF using the EURICA setup, revealed the presence of an isomeric transition at 1914(2)~keV with a proposed $J^\pi=(1/2^-)$ spin and parity  assignment~\cite{Escrig_PhysRevC.103.064328}.
This state differs by about 1~MeV from the shell model prediction, which assign this state as a prolate deformed state. A clear understanding on shape coexistence along the Co isotopic chain requires further investigation.


\subsection{Mn isotopes}
For the mid-shell isotopes Mn and V, spectroscopic information is very limited due to the high level density and their sensitivity to both the single-particle and collective degrees of freedom, which leads to a diversity of structures at low excitation energies. 
Studying these isotopes offers additional insights into shell evolution, while also providing robust means of testing the effectiveness of interactions developed for this particular region of the nuclear chart.

For \ts{63}Mn, the heaviest Mn isotope with a level scheme reported prior to the SEASTAR campaigns, the $7/2^-$, $9/2^-$, and $11/2^-$ excited states were measured and compared to shell model calculations using the LNPS interaction and various model spaces.  Based on such a comparison it was established that the calculation with the model space consisting of the $fp$ shell for protons and the $pf_{5/2}g_{9/2}d_{5/2}$ orbitals for neutrons, provided the best agreement with the experimental data~\cite{Bauger_PhysRevC.93.014313}.
Furthermore, the calculations indicated that the low-lying structure of \ts{63}Mn presents a significant mixing of the proton configurations, unlike in the simple single-particle approximation, where the structure would be dominated by a three $f_{7/2}$ proton holes configuration.
This necessity to include the $\nu g_{9/2}$ and $\nu d_{5/2}$ in lighter Mn isotopes had been pointed out by previous spectroscopy studies~\cite{Valiente_PhysRevC.78.024302,Chiara_PhysRevC.82.054313}, as well as by measurements of quadrupole moments~\cite{Babcock_PLB_750_2015,Babcock_PLB_760_2016,Heylen_PhysRevC.94.054321}.

During the SEASTAR~1 campaign, the neutron-rich \ts{63,65,67}Mn isotopes were produced by the fragmentation of \ts{68}Fe via the \ts{68}Fe$(p,2pxn)^{67-x}$Mn reactions with $x=4,2,0$, respectively. Level schemes for the three isotopes were deduced based on the coincidence analysis and the systematics in the region~\cite{Liu_PhysLettB.784.392}. 
Similarly to the lighter isotopes, a sequence of $7/2^-$, $9/2^-$, and $11/2^-$  excited states on top of the suggested $(5/2^-)$ ground state was proposed. 

\begin{figure}[t!]
    \centering
    \includegraphics[angle=-90,width=\textwidth]{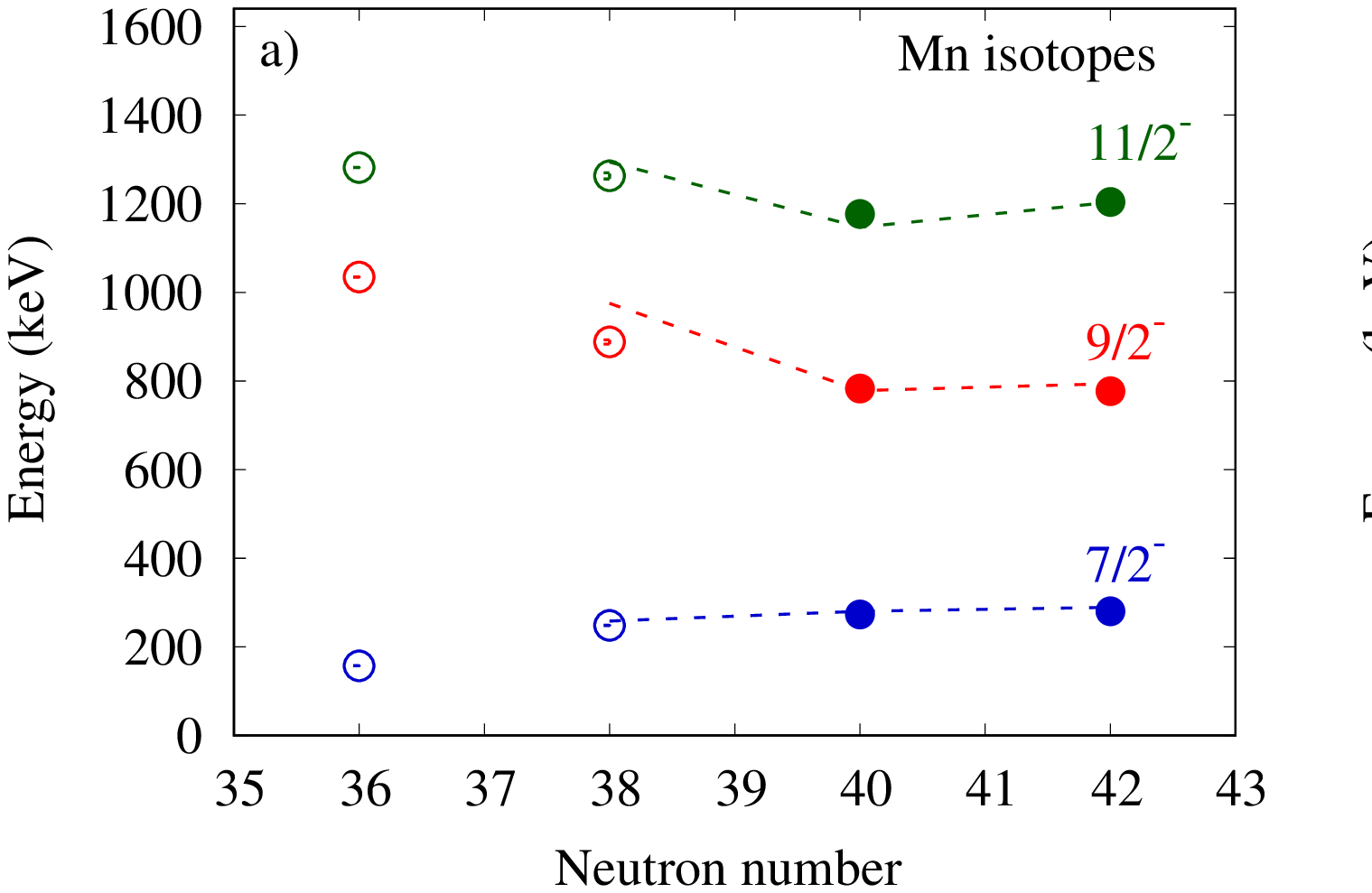}
    \caption{a) Evolution of low lying $7/2^-$, $9/2^-$ and $7/2^-$ states in Mn isotopes. Shell model calculations with the LNPS interaction are shown by the dashed lines. b) Energy of the low lying states in Mn isotpes as a function of $J(J+1)$. Adapted from~\cite{Liu_PhysLettB.784.392}.}
    \label{fig:MnEvolution}
\end{figure}

Figure~\ref{fig:MnEvolution}a) shows the evolution of  low-lying states  along the Mn isotopes. Shell model calculations using the LNPS interaction are shown by the dashed lines. 
It can be seen that the theoretical calculations give good account of the experimental results, in particular the decrease on the energies of the $9/2^-$ and $11/2^-$ states in \ts{65,67}Mn. 
The wave function of \ts{63,65,67}Mn results to be dominated by $4p-4h$ neutron excitations to the $g_{9/2}$ and $d_{5/2}$ orbitals beyond the $N=40$ shell gap. 
In addition, the calculation suggest that the $7/2^-$, $9/2^-$, and $11/2^-$ states belong to a rotational band with $K=5/2$.  

By plotting the excitation energy of the states as a function of $J(J+1)$, as shown in Fig.~\ref{fig:MnEvolution}b), the characteristics of a rotational band can be observed. 
For $N=40$ and $N=42$ a fairly linear relationship is present, showing a behaviour consistent with  textbook rotational bands. 
Below $N=40$, some degree of non-linearity is present, showing a possibility of a triaxial deformation, or a softness of the potential, as already proposed for \ts{63}Mn~\cite{Heylen_PhysRevC.94.054321},  indicating  an evolution towards strong coupling along the isotopic chain.

\subsection{V isotopes}

Spectroscopic information on V isotopes is even more scarce than for Mn isotopes, with \ts{57}V being the last even$-N$ isotope which had available information prior to the SEASTAR campaigns. 
For this isotope, a candidate for the $11/2^-$ state has been reported at 1.163~MeV based on \gammaray spectroscopy following multinucleon transfer reactions~\cite{Lunardi_AIP_2008}. Additionally, few states at excitation energies below 200~keV, as well as a group of states at around 1.7~MeV have been measured following the $\beta-$decay of \ts{57}Ti~\cite{Liddick_PhysRevC.72.054321_2005}. 
Shell model calculations employing the GXPF1A interaction in the $pf-$shell model space gave a moderate agreement with the experimental data  and suggested that the triplet of low-energy states could correspond to the $7/2^-$, $5/2^-$, and $3/2^-$ states, favoring the $7/2^-$ assignment for the ground state.

Spectroscopic information on \ts{59,61,63}V was obtained for the first time during the SEASTAR~3 campaign. 
Due to the limited acceptance of the SAMURAI spectrometer for these isotopes, low statistics were obtained and different reactions had to be  employed to populate excited states: 
For \ts{59}V the two-neutron knockout  reaction provided enough statistics to identify  four \gammaray transitions in the \gammaray spectrum. 
In the case of \ts{61}V,  neutron knockout and, with lower statistics, inelastic proton scattering  were employed revealing five transitions~\cite{Elekes_PhysRevC.106.064321}. 
Finally, in \ts{63}V mainly contributions from proton knockout and inelastic scattering  were employed and only two $\gamma$ rays were measured with a confidence level above 3$\sigma$~\cite{Juhasz_PhysRevC.103.064308}. 
Due to the limited statistics, only in the case of \ts{61}V a coincidence analysis was possible, establishing coincidences between some of the observed transitions~\cite{Elekes_PhysRevC.106.064321}. 

\begin{figure}[t!]
    \centering
    \includegraphics[angle=-90,width=\textwidth]{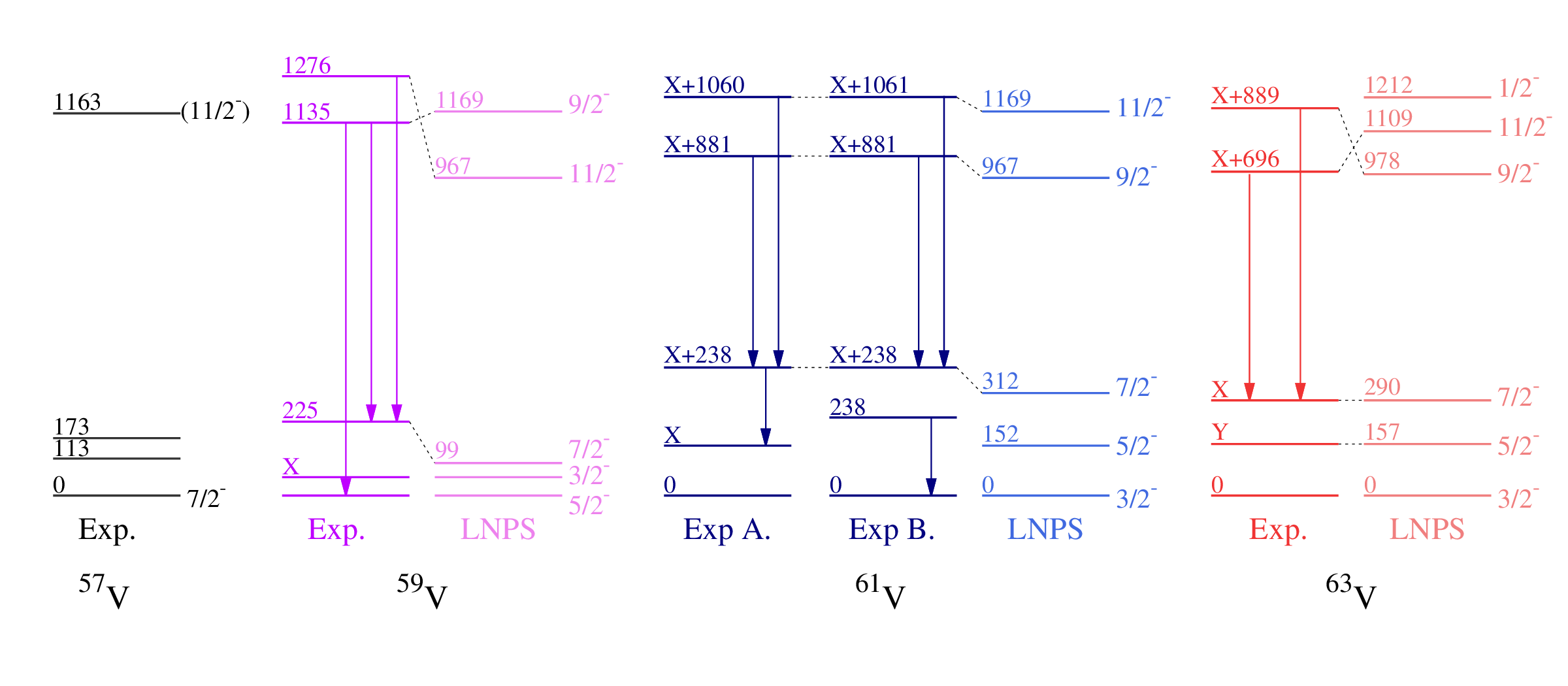}
    \caption{Experimental level schemes of neutron-rich \ts{57,59,61,63}V~\cite{Liddick_PhysRevC.72.054321_2005,Lunardi_AIP_2008,Elekes_PhysRevC.106.064321,Juhasz_PhysRevC.103.064308}. For \ts{59,61,63}V the experimental level schemes are compared with shell model calculations with the LNPS interaction.}
    \label{fig:VEvolution}
\end{figure}

Aiming to place the  observed \gammarays in a level scheme, shell model calculations with the LNPS interaction 
were performed.
The results of the calculations can be seen in Fig.~\ref{fig:VEvolution}.
In the three cases, two different sets of states are proposed, resembling the situation of \ts{57}V: At low excitation energies, below 400~keV, the $3/2^-$, $5/2^-$ and $7/2^-$ states are located. 
At higher excitation energies of around 1~MeV, the $9/2^-$ and $11/2^-$ states are predicted.
According to the same calculation, the ground state of \ts{61,63}V  is predicted to be the $3/2^-$, while for \ts{59}V, where the three levels lie within 100~keV, the $5/2^-$ is suggested as the ground state. 

For the case of \ts{59}V~\cite{Elekes_PhysRevC.106.064321}, the measured \gammaray energies are in better agreement with the decay of the $9/2^-$ and $7/2^-$ states. Based on this, and on the comparison between the theoretical branching ratios and the experimental intensities, three of the measured \gammarays were assigned to the level scheme, as shown in Fig.~\ref{fig:VEvolution}. 
A similar argument can be applied for the cases of \ts{61,63}V~\cite{Elekes_PhysRevC.106.064321,Juhasz_PhysRevC.103.064308}. In these cases the experimental $(p,p')$ cross sections were compared with coupled channels calculations assuming an axially symmetric rotor model. While predicted cross sections to the low-lying $3/2^-$, $5/2^-$ and $7/2^-$ states are much larger than the measured ones, a consistency was found for the population of the $9/2^-$ and $11/2^-$ states, strengthening the proposed spin and parity assignments and allowing to construct the level schemes shown in Fig.~\ref{fig:VEvolution}. It is noted that due to the high atomic background present in the experiment, it was not possible to measure low-energy transition, which leads to an ambiguity on the level scheme. Because of this, for \ts{61}V two possible levels schemes have been proposed, as shown in Fig.~\ref{fig:VEvolution}. 

The \ppp reaction on \ts{61,63}V was further exploited by the comparison of the measured cross sections with coupled channels calculations assuming only a quadrupole deformation, as well including an hexadecapole contribution. Although for \ts{61}V no firm conclusion was possible, for \ts{63}V deformation parameters  of $\beta_2=0.24(2)$ and $\beta_4=0.08(1)$  were obtained, indicating that this $N=40$ isotope belongs to the \ioi.


\subsection{Sc isotopes}

The Sc isotopes, with a single valence proton above the Ca isotopes, are 
an ideal testing ground for the 
study of  
neutron excitation in the $pf$ shell in the presence of an odd proton, as well as to the coupling of this valence proton with the Ca core. Indeed, in a shell-model picture, the coupling of the valence $f_{7/2}$ proton with the \stwop state of the Ca core would lead to multiplet of five states with spin and parity of $11/2^-$, $9/2^-$, $7/2^-$, $5/2^-$, and $3/2^-$.

For \ts{53}Sc ($N=32$), excited  states have been measured following multinucleon transfer reactions~\cite{Bhattacharyya_PhysRevC.79.014313}. 
Transitions at 2283, 345, and 2616~keV were measured to depopulate two states at 2.3 and 2.6~MeV.  The proposed level scheme showed a good agreement with shell model calculations in the $pf$ shell employing the UPF and GXPF1A interactions, and suggested that  these two levels correspond to the $9/2^-$ and $11/2^-$ states,  belonging  to  the multiplet arising from the coupling of the valence proton with the \ts{52}Ca core. 
A candidate for the $3/2^-$ state  at 2.1~MeV has been identified from $\beta-$decay studies~\cite{Crawford_APP_2009,Crawford_PhysRevC.82.014311} as well as from proton knockout~\cite{McDaniel_PhysRevC.81.024301}. 
Interestingly, in the proton knockout reaction the proposed $9/2^-$ and $11/2^-$ states were not observed, which is consistent with the proton knockout populating preferentially proton-hole states.  




The heaviest even-$N$ Sc isotope where spectroscopic information is available is \ts{55}Sc, located just above the recently established $N=34$ sub-shell closure. 
Low lying states on \ts{55}Sc were populated by proton knockout reaction and inelastic scattering, and based on coincidences analysis a level scheme was proposed and compared with  shell model calculations using the SDPF-MU interaction, and with {\em ab-initio} calculations. 
Based on the good agreement with the calculations, a candidate for the $3/2^-$ state was proposed at 695~keV. 
As no significant increase on the energy of this state was observed, unlike for the Ca isotopes,  it was concluded that there is no significant $N=34$ shell closure for Sc isotopes~\cite{Steppenbeck_PhysRevC.96.064310_2017}. 

No spectroscopic information is yet available for Sc isotopes towards $N=40$. 
Mass measurements for \ts{55-60}Sc have been recently reported showing a continuous declining trend towards $N=39$~\cite{Michimasa_PhysRevLett.125.122501}.
For \ts{61}Sc, only one proton above \ts{60}Ca, although its existence has been reported~\cite{Tarasov_PhysRevLett.121.022501}, no spectroscopic information is available. 

Detailed spectroscopy of \ts{55}Sc as well as first spectroscopy of \ts{57,59}Sc has also been achieved within the SEASTAR~3 campaign. Preliminary results of the analysis of \ts{55}Sc show agreement with the low-lying structure previously reported~\cite{Koseoglou_Sc_2020}, as well as a new transition at 1510~keV decaying directly to the ground state~\cite{Zidarova_2024}.
Further interpretation on these results, as well as the first spectroscopy of \ts{57,59}Sc will be discussed elsewhere~\cite{Zidarova_Sc}.


\section{Future perspectives}\label{sec:perspectives}

Further investigation on the $N=40$ \ioi can help us build a consistent view of shell evolution in the region, and to test the existing theoretical models, aiming to consolidate a unified description of these  nuclei. 

On the side of the even-even isotopes, the first spectroscopy of \ts{60}Ca would be a crucial milestone for the study of the $N=40$ \ioi. Only such a measurement can give direct evidence of this sub-shell closure for the Ca isotopes. Moreover, first spectroscopy of \ts{64}Ti, beyond $N=40$, can provide a strong benchmark for the thoretical models. 
However, as predicted by the shell model calculations with the  LNPS interaction, in spite of the fact that the \etwop for \ts{60}Ca is expected to increase with respect to the neighbouring \ts{62}Ti, a sizable shell gap at $N=40$ is unlikely. 
Measurement of this gap can only be obtained by mass measurements in this region, which are, therefore, of fundamental importance.
Furthermore, to study shell evolution beyond $N=40$, as well as the merging of the $N=40$ and $N=50$ Islands of Inversion, spectroscopic measurements on \ts{74,76}Fe are  necessary.

In addition, to have direct information on the collectivity around $N=40$ it is essential to measure transition probabilities. Lifetime measurements of \ts{56,58}Ti via lineshape analysis have been recently performed at the RIBF employing high-resolution Ge detectors  within the HiCARI project~\cite{Wimmer_RAPR_2021}.  These results can be extended in the more exotic \ts{60}Ti, as well as Cr and Fe isotopes.

For odd-$Z$ isotopes, of particular importance are the  measurements in Co isotopes, and specially the investigation on the low-lying $1/2^-$ states, which have indicated shape coexistence when going towards \ts{78}Ni. Deeper exploration of this phenomenon  requires the development of dedicated techniques to be able to measure the $0^+$ states in neighbouring even-even isotopes.

Upcoming results on the Sc isotopes will shed light not only on the shell evolution towards $N=40$ but also on the rapid changes observed for the $N=32$ and $N=34$ shell closures when adding a single valence proton. A comprehensive understanding of this region requires the development of shell model calculations which allow to include not only neutron excitations across $N=40$ towards the $g_{9/2}$ and $d_{5/2}$ orbitals, but also the contribution from the excitations of protons from the $sd$ shell. Such calculations are of particular importance in understanding the states observed following proton knockout reactions.

While some of these measurements are experimentally achievable at the RIBF with the current detectors, many others require significant upgrades both from the accelerator side, and  the detection systems. The planned RIBF upgrade will achieve unprecedented beam intensities~\cite{ribf_update}, which will in turn call for  improvements on the detection systems. In particular, for \gammaray spectroscopy, the development of the HYPATIA array will contribute significantly  to spectroscopic studies of very neutron-rich nuclei, where an improved efficiency is necessary to effectively use the available beam time~\cite{hypatia_web}. 
A review on the  detector developments for next-generation quasi-free scattering can be found elsewhere~\cite{Devices_PTEP}.


\section{Summary}\label{sec:summary}

Studies in shell evolution for exotic isotopes far from the valley of stability have revealed the appearance of new shell closures as well as the disappearance of some others.
In particular for $N=40$, which is a sub-shell closure arising  from the harmonic oscillator potential, an \iois has been established, where deformed intruder configurations overtake the contribution of  spherical configurations in the wave function. 

Within the SEASTAR project, the high intensity beams provided by the RIBF  have been efficiently exploited by the use of quasi-free reaction studies. The combination of the liquid hydrogen target of MINOS with the high-efficiency \gammaray detector arrays DALI2 and \dali have been crucial to study shell evolution around the $N=40$ \iois and its possible extension towards $N=50$.

A variety of results, particularly related to \gammaray spectroscopy have been discussed. Measurements on \etwop and \efourp  of even-even isotopes, as well as first spectroscopic studies of odd$-$Z isotopes have considerably increased the experimental information in this region. Likewise, inclusive and exclusive cross sections for direct reactions have been exploited to characterize the nature of the observed excited states. 

Shell model calculations employing the LNPS interaction in a model space consisting of the   $pf$ shell for protons and the $1p_{3/2}$, $1p_{1/2}$, $0f_{5/2}$, $0g_{9/2}$, and $1d_{5/2}$ orbitals  for neutrons has successfully reproduce the experimental data and has demonstrated a high prediction power regarding shell evolution in this region.  These calculations also suggest exciting possibilities, such as the merging of the $N=40$ and $N=50$ \ioi and shape coexistence.

Thanks to the upcoming upgrades on the RIBF accelerators and detection systems, as well as the next generation facilities FRIB and FAIR, further experimental  studies on exotic isotopes will be within reach, which will help us understand the mechanisms governing shell evolution and the fundamental nuclear force. 


\section*{Acknowledgment}

I would  like to thank all the speakers of the session devoted the the $N=40$ Island of Inversion during the symposium on \say{Direct reactions and spectroscopy with hydrogen targets: past 10 years at the RIBF and future prospects} held at York, UK in August 2023. 
I would like to specially acknowledge Silvia M. Lenzi, R. Taniuchi and S.~Chen for the careful reading of this manuscript and the many useful comments and suggestions.

\let\doi\relax
\bibliographystyle{ptephy}
\bibliography{ReviewRef}

\appendix

\end{document}